\documentclass[a4paper]{VisionStyle}
\usepackage{epsfig}

\begin{document}

\newcommand{\U}{{\sl Uhuru}}
\newcommand{\E}{{\sl Einstein}}
\newcommand{\A}{{\sl ASCA}}
\newcommand{\R}{{\sl ROSAT}}
\newcommand{\Y}{{\sl Yohkoh}}
\newcommand{\Ch}{{\sl Chandra}}
\newcommand{\XMM}{{\sl XMM-Newton}}
\newcommand{\HST}{{\sl Hubble Space Telescope}}

\newcommand{\Fig}{{\sl FIG}}
\newcommand{\Ref}{{\sl REF}}
\newcommand{\Tab}{{\sl TABLE}}

\title{X-rays from star-forming regions :
New objects, new physics, new visions}

\author{T.\,Montmerle\inst{1} \and N.\,Grosso\inst{2} \and 
  E.D.\,Feigelson\inst{3} \and L. Townsley\inst{3} } 

\institute{
  Service d'Astrophysique, CEA Saclay, France 
\and 
  Max-Planck-Institut f\"ur extraterrestrische Physik, Garching, Germany
\and
  Dept. of Astronomy \& Astrophysics, Penn State University, USA
   }

\maketitle 

\begin{abstract}

Star-forming regions have been the targets of X-ray observations since
the dawn of satellite X-ray astronomy.  The increase in sensitivity
and/or spatial resolution offered by \XMM\ and \Ch\ allows a
dramatic improvement, both qualitative and quantitative, on our
knowledge of high-energy phenomena in these regions and the underlying
physical processes.  We summarize here some recent developments:  the
Orion Nebula Cluster and its 1000+ stellar X-ray sources; Herbig-Haro
objects and their high-speed shocks; protostars, brown dwarfs and their
unusual magnetic activity; and the discovery of diffuse X-ray emission
from HII regions, presumably related to strong winds from massive
stars.  The role that future X-ray missions may play in the field 
is already starting to be visible.

\end{abstract}

\section{Introduction: an old new topic} 

\begin{figure}[ht]
  \begin{center}
    \epsfig{file=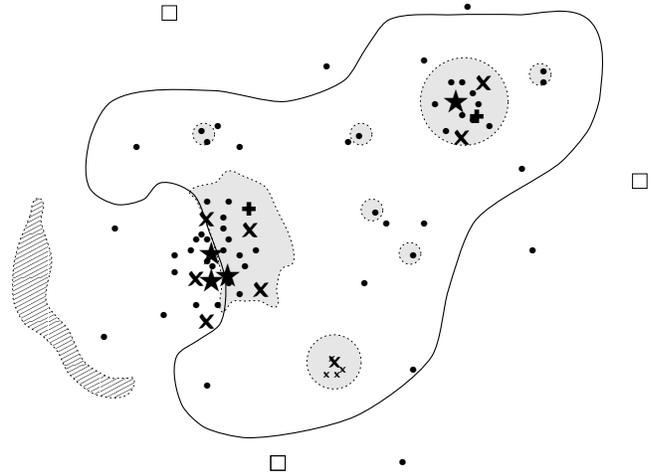, width=8.5cm}
  \end{center}
\caption{Diagram of the expected X-ray components from a giant
molecular cloud with a blister H{\sc II} region, embedded young star
clusters and distributed star formation.  
Symbols: {\Large\bf $\star$} =
OB stars; {\bf $\times$} = Herbig Ae/Be stars; $\bullet$ = T Tauri
stars; {\bf +} = protostars; {\it squares} = X-ray binary system.  The hatched
region outside of the cloud represents a supernova remnant, and shaded
regions within the cloud represent partially ionized X-ray dissociation
regions (see Feigelson 2001).
}
\label{fig:sfr_diagram}
\end{figure}

Star-forming regions have been known to be associated with X-ray
emission for over 30 years, beginning with the discovery of an
extended source coincident with the Orion Nebula (M42) using the
\U\, {\sl ANS} and {\sl SAS-3} satellites (Den Boggende et al.  1978, Bradt
\& Kelley 1979). Possibilities to produce X-rays from star
forming regions include magnetic activity from lower
mass pre-main sequence stars; thermalization of the high velocity winds
of higher mass OB stars, either close to the star or where at a wind
termination shock; and supernova remnants from past generations of OB
stars (Figure \ref{fig:sfr_diagram}). It took the advent of the first
{\it imaging} X-ray satellite, the \E\ Observatory, to realize that 
stars were the ``true'' X-ray emitters in the Orion Nebula.
Progress was rapid from the start, owing to the large field-of-view, 1
sq.  deg.  or more for satellites such as \E\ and \R, allowing to
detect dozens of stellar sources in a single exposure of nearby star
forming regions (see review by Feigelson \& Montmerle 1999, henceforth FM).

\begin{figure}[ht]
  \begin{center}
    \epsfig{file=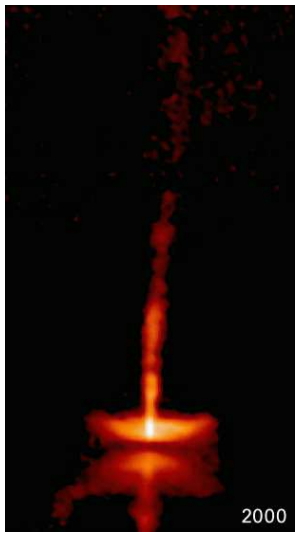, height=7cm}
  \end{center}
\caption{HST WFC2 optical view of the YSO HH30 in Taurus in 2000. The 
circumstellar,
flared disk, of radius $\sim 200$ AU, is seen almost edge-on. A bright,
collimated jet is expelled from the central regions.}  
\label{fig:HH30}
\end{figure}

In parallel with the development of X-ray satellites, major progress was
being achieved in sensitive solid-state detectors for ground-based
telescopes in the IR and mm ranges.  This led to the discovery of {\it
circumstellar material}, accretion disks and envelopes around young stars
and protostars (see, e.g., Andr\'e \& Montmerle 1994).  High-angular
resolution imaging, by, e.g., the \HST\ and mm interferometers, showed that
young stars and protostars (``Young Stellar Objects'':
YSOs) lose mass in the form of jets and outflows, collimated perpendicular
to the circumstellar disks.  Thus accretion and ejection are currently
viewed as correlated phenomena, somewhat paradoxically required to build up
a star. Fig.~\ref{fig:HH30} gives a spectacular example, obtained with the
\HST, of a disk-jet system associated with a very young star.

\begin{figure}[!ht]
  \begin{center}
    \epsfig{file=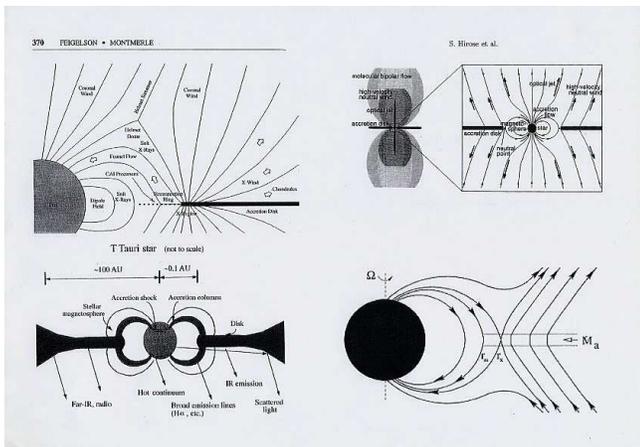, width=8.5cm}
  \end{center}
\caption{Various star-disk magnetic configurations, taken from the 
literature. Top left: Shu et al. (1997). Top right: Hirose et al. (1997).
 Bottom left: Hartmann (2000). Bottom right: Ferreira et al. (2000).}  
\label{fig:MagConf}
\end{figure}

It is widely believed that magnetic fields are required to explain, and
perhaps even cause, the ejection of material, from the disk, and/or
from the central, growing star (e.g., K\"onigl \& Pudritz 2000).
Fig.~\ref{fig:MagConf} illustrates various star-disk magnetic
configurations from the literature. Recent calculations indicate that
YSO X-rays which, as on the Sun, are produced in violent magnetic
reconnection events, may be the principal ionization source at the base
of these jets (Shang et al. 2002).  We thus see that in YSOs X-rays,
circumstellar matter, and magnetic fields are somehow intimately
interrelated.

\section{Activity properties: low-mass young stars}

The X-ray properties of T Tauri stars and protostars are now well-known
(FM, Imanishi et al. 2001a, Feigelson et al. 2002a), and can be
summarized as follows:

$\bullet$ {\it X-ray luminosities}: $L_X < 10^{28}$
to $10^{31}-10^{32}$ erg s$^{-1}$; X-ray to bolometric
luminosity ratio $L_X/L_{bol} \approx 10^{-4}$. 

$\bullet$ {\it Emission mechanism}: Bremsstrahlung emission from plasmas
with $T_X \sim 1 - 10$ keV dominate the X-ray spectra.  Emission lines
from highly ionized species are clearly present when high-quality
spectra are obtained.

\begin{figure}[ht]
  \begin{center}
    \epsfig{file=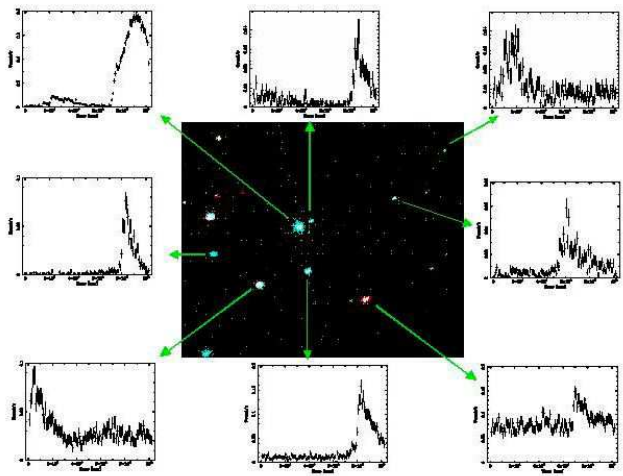, width=8.5cm}
  \end{center}
\caption{X-ray flaring activity in the $\rho$~Oph cloud core region,
as seen during a 100 ksec exposure with \Ch\ (Imanishi et al. 2001a): the
brightest source is YLW16A (circled in Fig.~\ref{fig:EinsteinChandra}).
}  
\label{fig:Babystars}
\end{figure}

$\bullet$ {\it Ubiquitous flaring activity} on
timescales of minutes to days, but more typically $\sim$ a few hours, with
amplitudes peak/quiescent flux up to $\sim 100$ (Figure \ref{fig:Babystars}).

$\bullet$ {\it Extinction}: Soft X-ray absorption ranges from
negligible to very high for embedded objects like protostars where
column densities can be $N_H \sim 10^{23}$ cm$^{-2}$, corresponding to
visual absorptions around $A_V \sim 100$.

In addition to X-ray flares, non-thermal {\it radio} emission testifies to
the existence of $\sim$ MeV electrons (accelerated in flares),
radiating via the gyrosynchtrotron mechanism, based on flarelike
variability, and/or on significant polarization.  In contrast,
thermal radio emission is attributed to ionized material, either strong
winds in the case of T Tauri stars, or hot jets in the case of protostars
(FM).

\begin{figure}[ht]
  \begin{center}
    \epsfig{file=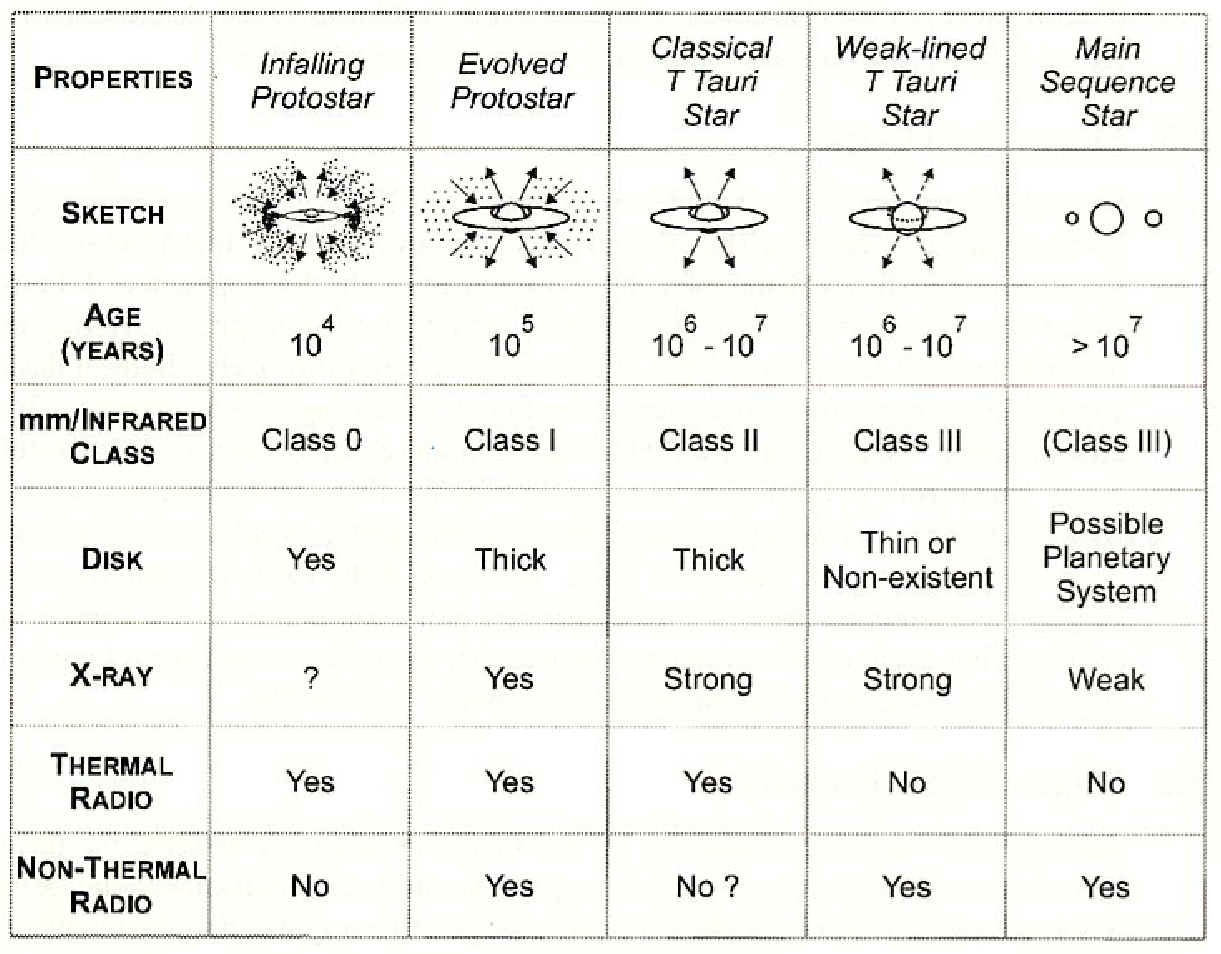, width=9cm}
  \end{center}
\caption{YSO properties and evolutionary classes in a nutshell 
(see FM for details).}  
\label{fig:Table 1}
\end{figure}

\begin{figure*}[ht]
  \begin{center}
    \epsfig{file=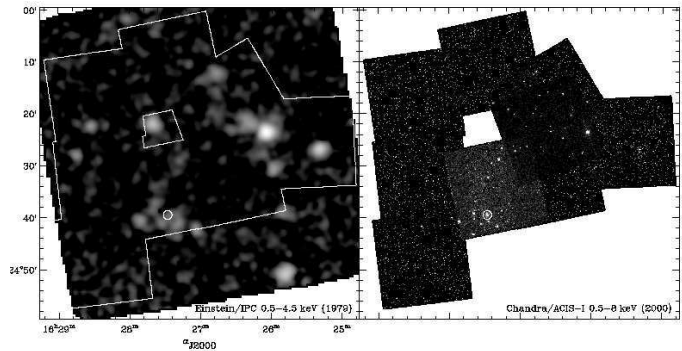,angle=90,height=7cm}
  \end{center}
\caption{\E-IPC image (1979) vs. \Ch-ACIS (2000) mosaic of 
the $\rho$~Oph cloud core (from archival data). A small circle 
is drawn around the Class I protostar YLW16A.}  
\label{fig:EinsteinChandra}
\end{figure*}

Fig.~\ref{fig:Table 1} summarizes the observational properties of
YSOs, according to their evolutionary state, from very young ($\approx
10^4$ yrs), envelope-dominated ``Class 0'' protostars, to evolved ($\approx
10^6-10^7$ yrs) ``weak-lined'' (i.e., diskless) ``Class III'' T Tauri
stars.  Note that the most recent results from \Ch\ and \XMM\
have not significantly modified the summary shown in Fig.~\ref{fig:Table 1}.

The case of protostars deserves some comment.  Although the detection of X-ray
emission from Class 0 protostars has not been confirmed yet, the X-ray emission
from Class I protostars (age $\approx 10^5$ yrs) is now widespread, following
their announced discovery in the late 90's with \A\ and \R\ (Koyama et al.
1996, Grosso et al.  1997).  However, going back to the \E\ observations of
Montmerle et al.  (1983), it turns out that a Class I protostar had already been
detected (see discussion in Grosso 2001) !  The historical point, of course, is
that protostars were not known as such at the time, for lack of adequate IR and
mm data.  Fig.~\ref{fig:EinsteinChandra} shows, next to
each other, the two images of the $\rho$ Oph cloud from \E\ and from \Ch\ (from
archival data), which are separated by 21 years:  the vastly improved angular
resolution of \Ch\ allows accurate identifications of the X-ray sources, but the
YLW16A protostar is nevertheless visible on the corresponding {\sl Einstein}
image (circle).  Fig.~\ref{fig:Babystars} shows the flarelike variability of
X-ray sources in the same region, as observed in a $\sim 100$ ksec exposure by
\Ch\ (Imanishi et al.  2001a).

\section{New satellites, new objects}

With the advent of a new generation of wide-band ($\sim 0.2 - 10$ keV),
sensitive X-ray satellites, i.e., \Ch\ and \XMM, new problems can be
addressed, new objects can be detected, and new processes can be
studied.  We illustrate this by briefly reviewing representative recent
results on X-ray emission from YSOs and star-forming regions.

\subsection{The Orion Nebula Cluster}

\begin{figure*}[ht]
  \begin{center}
    \epsfig{file=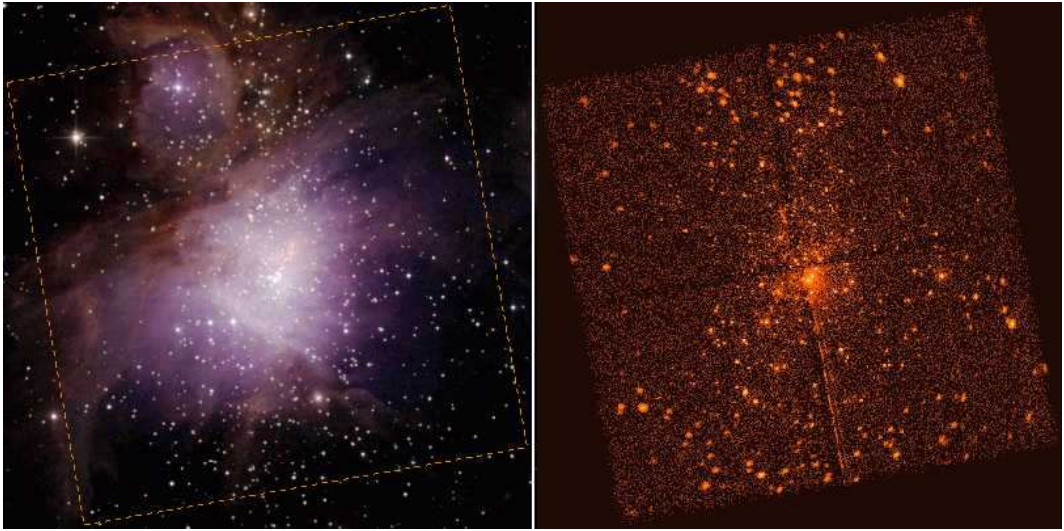,height=7cm}
  \end{center}
\caption{2MASS (left) vs. \Ch-ACIS (right) images of the Orion Nebula 
Cluster (Garmire et al.
2000). Note the excellent correspondence between the near-IR and X-ray 
stellar sources.}  
\label{fig:ONC_2MASSChandra}
\end{figure*}

Illuminated by the ``Trapezium'' OB stars, the Orion Nebula hosts
a cluster of 2000+ lower-mass stars called the Orion Nebula
Cluster (ONC; Hillenbrand 1997, O'Dell 2000).  
Fig.~\ref{fig:ONC_2MASSChandra} (right) shows the first of two
$\sim 40$ ksec exposures (Garmire et al. 2000).  In the $17' \times
17'$ field-of-view of the ACIS-I CCD detector lie over 1000 ONC stars,
most detectable in the near-IR which can penetrate moderately large
column densities  (Fig.~\ref{fig:ONC_2MASSChandra}, left).  This
ACIS image of the ONC is so far the richest X-ray image ever obtained
and also contains a wealth of temporal and spectral information.

A variety of findings have emerged so far from the \Ch\ Orion
Nebula studies (Feigelson et al. 2002a \& b). Several dozen new YSOs,
some in the lightly absorbed ONC and others deeply embedded in the
molecular cores, are discovered.  X-ray emission is found to be
surprisingly hot ($T_X > 5$ keV) in a fraction of the low mass pre-main
sequence stars, characteristically hotter than those seen in older main
sequence stars, and certainly far above the temperatures seen in
quiescent solar-like coronae. Curiously, low-mass X-ray emission
appears to depend more on stellar mass or luminosity than on rotation,
as it does in main sequence stars. The Orion image also provides the largest
sample of X-ray detected pre-main sequence brown dwarfs (see below).
The majority of sources show prominent flares: solar-mass stars produce
flares with $L_{X,peak} > 10^{29}$ erg s$^{-1}$ roughly once every few
days, each lasting 2 to $>12$ hours.  Surprisingly, the $\sim 30$
M$_\odot$ O9.5 star $\theta^2$A Ori exhibited rapid variations also.
This is inconsistent with the standard model of production in a myriad
small wind shocks and probably requires some near-surface magnetic
processes.

\subsection{Herbig-Haro objects}

During the early fifties, Guillermo Haro and George Herbig
independently discovered, associated with dark clouds, tiny luminous
spots having ``nebular'' optical spectra showing sharp emission lines
and no continuum. Dubbed ``Her\-big-Haro objects'', over 500 of these
nebulae are now known in star-forming regions (Reipurth \& Bally
2001).  They are interpreted as ambient cloud material shocked by
high-speed ($\sim 200-300$ km s$^{-1}$) jets powered by protostars and
young T Tauri stars (as shown on Fig.~\ref{fig:HH30}).

Although an early \E\ study reported the discovery of X-ray
emission from HH objects (Pravdo \& Marshall 1981), subsequent
observations did not confirm it.  It is only recently that both
\Ch\ and \XMM\ have allowed to convincingly detect X-rays from HH
objects in Orion and Taurus.  Using \Ch, Pravdo et al.  (2001) detected
a few X-ray photons from the tip of HH2 (part of the HH1 system) in
Orion, while Favata et al.  (2002), using \XMM, detected faint X-ray
emission from the blue-shifted jet of the L 1551 IRS 5 protostar in
Taurus.  In both cases the photons are soft, consistent with a
shock velocity $\sim 200$ km~s$^{-1}$. Many other HH objects are not
seen in X-ray images, presumably due to higher obscuration and/or lower
shock temperatures.

\subsection{Protostars}

Whereas only a small fraction of the nearest Class I protostars
were detected by \R\ and \A\ (FM), \Ch\ and \XMM\ now detect about 70
\% of the Class I protostars in nearby YSO clusters (Imanishi et al.
2001a; Getman et al. 2002).  The X-ray luminosities and time
variability characteristics seem in general very similar to those of T
Tauri stars, although protostars tend to have
characteristically harder spectra.  A remarkable case is the
Class I protostar YLW15 in the Ophiuchus cloud, which has been
detected on several occasions at both ``low'' and ``high'' states,
and displayed a $\sim 20$h quasiperiodic sequence of three flares
observed by \A\ (Tsuboi et al.  2000).  (Unfortunately, it was in its
``low'' state during \Ch\ and \XMM\ observations.) The case of the
youngest, Class 0 protostars is less clear, with tentative detections
in \Ch\ images of the OMC 2/3 clouds in Orion (Tsuboi et al.  2001)
but no further detection in several nearby clouds so far.

\subsection{Brown dwarfs}

At the lowest end of the stellar mass function, brown dwarfs are
especially interesting since their effective temperature is so low ($<
3000$ K) that their photospheres have a very low ionization degree.
X-ray emission from brown dwarfs was first discovered by \R\ in the
Chamaeleon and other nearby clouds (Neuh\"auser \& Comer\'on  1999).
Using \Ch, Feigelson et al.  (2002a) detected 30 very low mass objects
in the ONC, and smaller samples are seen in IC 348 and Ophiuchus
(Preibisch \& Zinnecker 2001, 2002; Imanishi et al. 2001b).  This now large
sample of pre-main sequence very low-mass objects ($0.02 < M < 0.1
M_\odot$) is characterized by strong surface magnetic activity very
similar to those of late-type stars, with $ L_X/L_{bol} \sim 10^{-3}$
at the saturation level, hard spectra, and flares.  In contrast, only
one old brown dwarf ($\sim$500\,Myr) has been identified by \Ch\ so far
while flaring (Rutledge et al.  2000).

\subsection{Diffuse emission from HII regions}

\begin{figure}[ht]
  \begin{center}
    \epsfig{file=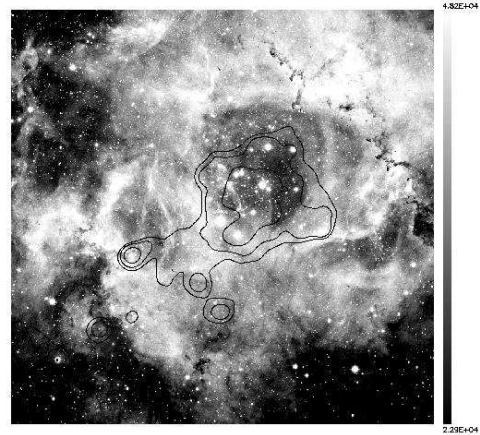,width=7.5cm}
  \end{center} 
\caption{Optical/X-ray image of the Rosette nebula: \Ch\-ACIS contours are 
shown overlaid on an 
H$\alpha$ image (kindly supplied 
by T.A.~Rector, B.~Wolpa, and M.~Hanna, AURA/NOAO/NSF).
About 350
point X-ray sources have been removed, leaving a weak, soft diffuse X-ray
emission (contours; Townsley et al. 2002, in
preparation).
} \label{fig:Rosette_diff} \end{figure}

Evidence for {\it diffuse} X-ray emission has been found for the first time in
HII regions by Townsley et al.  (2002, in preparation), i.e., in the Rosette
Nebula.  Powered by the OB association NGC 2244, this thick shell-like blister
H{\sc II} region lies on the edge of a giant molecular cloud at a distance of
1.4 kpc, and contains two early O stars plus several late O and early B stars.
The radio H{\sc II} region and visible band nebula exhibit a large central
cavity, $\sim$10~pc in diameter.  As depicted in Fig.~\ref{fig:Rosette_diff},
the new \Ch\ study shows that this cavity contains X-ray emitting plasma.

The high-resolution mirrors
of \Ch\ are critical for resolving the diffuse emission from the
$\sim$350 faint stellar sources. A spectral fit to the diffuse
emission obtained (after removal of the instrumental background, point
sources and detector charge transfer inefficiency effects) are
consistent with an absorbed Raymond-Smith plasma $N_H = 7 \times
10^{21}$~cm$^{-2}$, kT = 0.6~keV and $L_{X} \simeq 8 \times
10^{32}$~erg s$^{-1}$ in the $0.5-2$ keV band after correction for
absorption.  Based on the X-ray luminosity function (XLF) found by
\Ch\ in the ONC (see above), Townsley et al.  estimate that the
unresolved point source population (made up of faint, low-mass stars)
should not be contributing more than to 25 \% of this emission.  As
outlined below, this nebulous X-ray emission in the Rosette Nebula, and
an even more dramatic case found very recently in the M17 H{\sc II}
region, is best explained as the result of collisions and shocks
between OB stellar winds and the ambient cooler medium (Townsley et
al., in preparation).

\section{New physics}

\subsection{Coronal activity and star-disk interactions}

X-rays from pre-main sequence low-mass stars, as in the Sun and other
older late-type stars, are most easily interpreted as emission from
plasma heated and confined by multipolar magnetic fields rooted in the
stellar surface (FM).  The most convincing arguments are based on the
flaring activity.  However, it has proved difficult to establish the
properties of these flares with any confidence.  While X-ray flares
from T Tauri stars had often been modeled as uniform plasma in very
large ($l > R_\star$) magnetic loops undergoing radiative cooling, it
is possible that many flares arise in smaller, complex loop
arcades which reconnect and reheat the plasma on timescales of hours.
The inferred emitting volume is then considerably reduced compared to
that implied by pure radiative cooling (Favata et al. 2001).  In the
one case where a T Tauri star is sufficiently bright to give a good
high-resolution \Ch\ grating spectrum, TW Hya, the situation is even
more confusing with plasma densities inferred from helium-like triplet
lines (see Porquet et al. 2001) much higher than those seen in any normal 
stellar flare (Kastner et al. 2001).

The real issue, however, is whether magnetic configurations other than
solar may exist.  The younger YSOs (from Class 0 to Class II) are
surrounded by accretion disks, which are widely thought to harbor
magnetic fields, originally interstellar and dragged by gravitational
collapse, perhaps amplified by an intra-disk dynamo as well.  While the
exact configuration is still debated, all models agree that, in order
to provide a satisfactory framework to drive jets and outflows from
accretion, star-disk magnetic structures must exist
(Fig.~\ref{fig:MagConf}).  If some twisting of the field lines is
included, resulting for example from shear in the Keplerian disk or
fluctuations in the inner accretion disk, flares may occur (Shu et
al. 1997).  This situation has been numerically studied by Hayashi et
al.  (1996), assuming some finite resistivity for the field lines:  in
brief, powerful flares with X-ray emitting plasma and ``coronal mass
ejections'' are produced at the star-disk corotation radius.  This
situation may have already been observed in the triple X-ray flare of
YLW 15 (Tsuboi et al. 2000). Montmerle et al. (2000) propose that
three $\sim 20$h successive reconnection events occurred in a star-disk 
magnetic configuration, in which the protostellar rotation is very rapid and
not yet decelerated by interaction with the disk.

The X-ray emission from brown dwarfs also presents new challenges in
coronal physics. Despite their similarities with late-type stars, there
is clear evidence from H$\alpha$ surveys of older L- and T-type brown
dwarfs in the solar neighborhood that their chromospheric and coronal
activity drop at the stellar/sub-stellar boundary (e.g.  Gizis et al.
2000).  This is probably related to the fact that these very cool
objects ($T_{eff} < 3000$\,K) have essentially neutral atmospheres with
high electrical resistivity, in which the rapid decay of currents
prevents the buildup of magnetic free energy and therefore cannot
provide support for magnetically heated chromospheres and coronae.
However the very low-mass stars for which quiescent X-ray emission was
detected are all very young, and their atmosphere is still warm enough
to be partially ionized and therefore capable of sustaining electrical
currents.  The transient coronal plasma, which gives rise to the
observed X-ray flare in old very low mass stars, could be created by
buoyant flux tubes that are generated in the interior and rise rapidly
through the atmosphere, dissipating their associated currents in the
upper atmospheric layers (Fleming et al.  2000; Mohanty et al.  2002).

\subsection{Which dynamo ?}

Another significant puzzle is the weakness or absence of a statistical
correlation between X-ray emission and stellar rotation in T Tauri
stars compared to the very stong relation seen in main sequence stars.
The problem is most dramatic in the new \Ch\ ONC dataset where over 200
well-characterized pre-main sequence stars can be placed on the
$L_X-P_{rot}$ diagram (Feigelson et al., in prepration).  A random
scatter diagram rather than a tight correlation is seen : {\it mass
rather than rotation appears to be principal determinant of X-ray
activity}.

The reasons for this are not at all clear.  One possibility is, as
discussed just above, that the X-ray flares arise from disk-related
magnetic fields rather than solar-like fields.  But X-ray properties
seem remarkably insensitive to the presence or absence of an infrared
disk.  Another possibility is that we are seeing the effects of a
different magnetic dynamo in the interiors of YSOs.  Main sequence
stellar activity is attributed to an $\alpha-\Omega$ dynamo where the
fields are generated at the boundary between the
convective and radiative zones.  But T Tauri stars are nearly fully
convective, and their magnetic fields may be generated entirely within
the convective zone ($\alpha^2$ or turbulent dynamos).  Here the
relationship between surface rotation and magnetic fields may break
down (Kuker \& Stix 2002).

\subsection{Disk irradiation by X-rays and particles}

The question of star-disk vs.  star-star magnetic topology is of
importance not only for nagnetohydrodynamical aspects of flare and
outflow physics, but also for key issues relevant to the earliest
stages of the evolution of protoplanetary disks.  We outline three
aspects of this complex issue.

\begin{figure}[ht]
  \begin{center}
    \epsfig{file=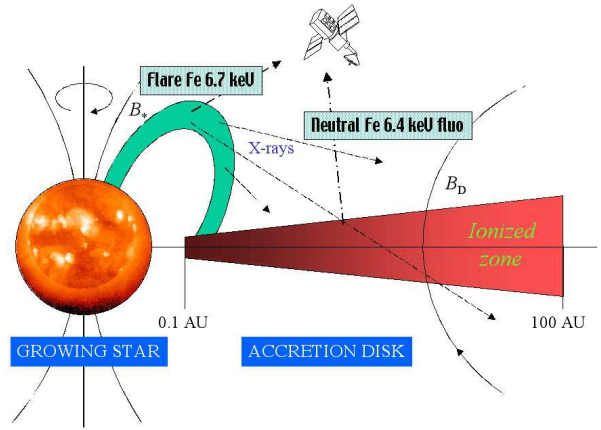,width=8.5cm}
  \end{center}
\caption{X-ray irradiation of a circumstellar disk. A high-latitude
stellar X-ray
flare (emitting the ``hot'' Fe 6.7 keV line) shines over the disk, generating
the ``cold'', fluorescence Fe 6.4 keV line.}  
\label{fig:X-irrad-fluo}
\end{figure}

$\bullet${\it X-ray ionization effects}.  X-ray flares must impact some
part of the circumstellar disks and partially ionize the largely
neutral material.  The uncertain geometry of the X-ray emitting
magnetic structures controls the size of the portion of the disk that
will be irradiated.  With reasonable X-ray luminosities and
optimistic geometries, Igea \& Glassgold (1999) have shown that X-ray
ionization (generally low, $n_e/n_H < 10^{-7}$) will induce the
MHD magneto-rotational (Balbus-Hawley) instability, which in turn
stimulates turbulent viscosity and promotes accretion onto the young
star.  Ionization of the outer layers of the disk may be critical
for coupling the Keplerian rotation of disk material to the collimated
outflows (Ouyed \& Pudritz 1999).

$\bullet${\it Spallation by MeV particles}.  As already mentioned, in addition
to voluminous X-ray results on YSO flaring, radio continuum studies
show that, as in solar flares, MeV electrons are copiously produced in
(at least some) YSO flares (FM).  MeV protons are also probably
accelerated in these violent magnetic reconnection events.  An estimate
of this effect for $\simeq 1$ M$_\odot$ YSO analogs of the young Sun in
the \Ch\ observation of the ONC indicates a $10^5$-fold enhancement in
energetic protons compared to contemporary levels (Feigelson et al.
2002b).

This result may have profound implications for a long-standing
puzzle in the meteoritic record of conditions in the protoplanetary
disk around the early Sun (Feigelson et al. 2002b).  Anomalously high
levels of daughter products of several radioactive nuclei with short
lifetimes (e.g. $^{41}$Ca, $^{26}$Al, $^{53}$Mn) are found in
calcium-aluminum-rich inclusions of carbonaceous chondrites such as the
Allende and Murchison meteorites.  These have traditionally been
attributed to the injection of recently synthesized nuclides by a
supernova remnant near the molecular cloud that formed our solar
system.  However, the $10^5$ elevation in MeV protons inferred from the
\Ch\ data is just the level found in recent calculation of radioisotope
production by spallation of disk solids.  X-ray astronomy thus provides
a novel bridge for the study of the origins of our solar system.

$\bullet${\it The neutral Fe 6.4 keV fluorescence line}.  In a fashion
inspired from AGN studies (e.g.  Nayakshin \& Kazanas 2002), one can
look for neutral Fe K$\alpha$ line emission at 6.4\,keV, in addition to
the 6.7\,keV emission line from the flare plasma, resulting from the
fluorescence of the accretion disk induced by the flare X-ray
irradiation.  A strong double-line feature may imply that the X-ray
emitting plasma is located high above the disk in a ``lamppost''
configuration which would lead to efficient ionization and spallation
in the disk (see Fig.~\ref{fig:X-irrad-fluo}). This double-Fe line spectral
signature has been seen to date during flares of two protostellar X-ray
sources (Koyama et al.  1996, Imanishi et al.  2001a).  In principle,
if sufficient signal over many hours could be obtained, the X-ray/disk
geometry could be revealed by reverberation mapping.

$\bullet${\it Disk chemistry}.  While the chemistry of YSOs disks had been
long studied in the context of the solar nebula and formation of our
planetary system, advanced in millimeter and infrared imaging and
spectroscopy is providing many new insights.  Among them are hints of
chemical effects attributable to X-ray irradiation.  Calculations
indicate that X-rays should significant heat the gas above the dust
temperature by hundreds of degrees in the outer layers of the disk,
altering the chemistry and expected molecular spectroscopic signature
(Glassgold \& Najita 2001).  Kastner et al.\ (1997) attributed the high
CN/HCN ratio and high HCO$^+$ abundance in the disk of TW~Hya to X-ray
illumination.  Similarly, the outer disk of LkH$\alpha$ 15 shows a
$1-3$ order of magnitude excess of HCN and CN compared to chemical
models, which might be explained by X-ray ionization and dissociation
which promotes to transformation of CO into cynanides (Aikawa et al.
2002).

\subsection{Shock physics}

Up to now, X-ray emission in young stellar objects has been discussed
principally in the context of magnetic activity.  However, as more
sensitive observations become possible, new, weaker processes become
accessible, like interstellar shocks (Fig \ref{fig:sfr_diagram}).
Since X-ray emission results from the thermalization of shocked gas
flowing at velocity $v_s$, the approximate temperature $T_X$ of a
strong, adiabatic shock, can be found from  $T_X= (3/16k) \mu m_p v_s^2$:  
$T_X \sim 10^6$ K for $v_s
\sim 200 - 300$ km~s$^{-1}$  and $T_X \sim 10^7$ K for $v_s \sim 1000$
km~s$^{-1}$.  We thus expect soft X-ray production associated with
accretion flows and Herbig-Haro outflows from low-mass YSOs, and hard
X-rays associated with the winds of massive stars.  Study of these
processes presents considerable observational difficulties, first due
to the absorption of soft X-rays by line-of-sight material and second
due to the difficulty in resolving truly diffuse emission from
unresolved stellar emission in rich YSO clusters.

However, three lines of recent evidence point to such processes.  First,
according to Kastner et al.  (2002), high-resolution \Ch/HETGS spectroscopy of
the nearest and brightest classical T Tauri star, TW~Hya, indicates the dominant
plasma is cooler ($T_X = 3 \times 10^6$ K) and denser ($n_e \approx 10^{13}$
cm$^{-3}$) than usual flare plasmas, and further shows large abundance
anomalies.  Such characteristics are consistent with a shock from mass being
accreted at a velocity $\approx 200$ km~s$^{-1}$ and at a relatively high rate
$\dot{M} \sim 10^{-8} M_\odot$\ yr$^{-1}$, although the abundance anomalies are
more consistent with a surface flare model.

Second, the detection of soft X-rays from HH-2 (Pravdo et al.
2001) is a very promising of shock emission.  It is interpreted as hot spots at
$\sim 10^6$ K resulting from the collision of the $\sim 250$ km~s$^{-1}$ jet
with the ambient interstellar medium.  Discussing the case of L1551, Favata et
al.  (2002) raise interesting astrophysical issues such as X-ray ionization on
ambient cold gas, leading to an enhanced coupling of matter with
magnetic fields.

The third case of energetic shock processes in star formation regions
is clearly indicated by the discovery of diffuse X-rays in rich OB
associations like Rosette and M 17 (Townsley et al., in preparation).
Several percent of the radiative luminosity in the earliest O stars is
converted into wind mechanical luminosity with mass-loss ($\dot{M} >
10^{-6} M_\odot $~yr$^{-1}$, $v_w \simeq 1000 - 2500$~km~s$^{-1}$) and
$L_W \sim \frac{1}{2} \dot{M} v_w^2 \sim 10^{36}-10^{37}$ erg
s$^{-1}$.  A naive estimate shows that the thermalization of the wind
should yield post-shock temperatures $\sim$1--5~keV (Weaver et al.
1997).  A closer look at the interaction between the winds and the
surrounding HII region shows however that the problem is quite complex,
and is still not resolved theoretically 35 years after the discovery of
stellar winds from early-type stars.  For example, it has long been
recognized that the Rosette Nebula is much smaller than expected from a
wind-blown bubble with an age of a few $10^6$ yrs (Oey 1996; Bruhweiler
et al.\ 2002).

Several physical effects may complicate the thermal structure of the
wind shock and account for the age-size discrepancy:  dissipation
in a turbulent mixing layer between an ionization-bounded HII layer and
a hot shocked stellar wind (Kahn \& Breitschwerdt 1990); non-local
conductive electron transport, leading to nonthermal hard X-ray emission
(Dorland \& Montmerle 1987); and mass loading of the stellar winds with
interstellar material, heating the flow near the stars and weakening
the terminal shock (Pittard, Hartquist, \& Dyson 2001 and refs
therein). Hydrodynamical calculations of X-ray emission for two
cases of OB wind-blown bubbles (Strickland \&
Stevens 1998; Cant\'o, et al. 2000) predict emission to several
keV in the parsec-scale wind collision zones between the principal O
stars in the association.

So while we have long been expecting diffuse X-ray emission from HII
regions, the possible interpretations of the data at hand are diverse,
and the work in progress by Townsley et al.  should shed a fresh light
on the new physics of the interaction of massive stars with their
environment.

\section{Conclusions}

As in other fields of X-ray astronomy, \Ch\ and \XMM\ have opened a new
era for the study of star-forming regions and the early stages of
stellar evolution.  This era is only starting, since at the time of
this Conference only a limited amount of results are available.
However, the selection we have presented above already demonstrate the
new possibilities:

$\bullet$ In dense young clusters, like the ONC, several
hundreds of low-mass stars can be detected and identified
simultaneously in one image, from massive OB stars down to substellar objects
which will evolve into brown dwarfs;

$\bullet$ Protostars are detected with a high efficiency. Most
detected protostars are Class I (evolved, age $\sim 10^5$ yrs), along
with two candidate Class 0 (young, age $\sim 10^4$ yrs);

$\bullet$ X-ray flares testify to the intense magnetic activity
(stellar surface or star-disk reconnection events) of YSOs, and provide
a unique probe of magnetic fields associated with the
``central engine" powering jets and outflows;

$\bullet$ X-ray activity, down to very low levels, has an important 
impact (via irradiation processes) on physical conditions in 
protoplanetary disks, in particular ionization and coupling of material 
with magnetic fields, and by inference on the early solar system where
products of internal spallation reactions may be recorded today in 
anomalous meteoritic abundances;

$\bullet$ The increase in sensitivity and/or angular resolution has
enlarged the number of observable mechanisms for YSO X-ray emission, to
include shocks (both from accretion in T Tauri stars, and from jets in
Herbig-Haro objects), and shock-related pc-scale diffuse emission
processes from stellar winds of massive stars in HII regions.

At the same time, the limitations of \Ch\ and \XMM\ already begin to be
visible:  for instance, high-resolution spectroscopy at 6.4 keV
($\Delta E << 100$ eV), combined with a higher throughput to allow time
resolution, is necessary to use ``reverberation
mapping'' to probe the circumstellar disks of YSOs.  So, even as \Ch\ and
\XMM\ results on star-forming regions keep coming, it is already time
to follow the next-generation X-ray projects like {\sl Astro-E II}, {\sl XEUS} 
and {\sl Constellation-X} !

\end{document}